\documentclass[preprint]{JASA}
\begin{document}

\title[JASA]{A Self-Supervised Approach for Minimal-Annotation Hydroacoustic Data Exploration}

\author{Pierre-Yves Raumer}
\affiliation{Laboratoire de Géologie, Ecole Normale Supérieure/CNRS UMR 8538, PSL Research University, Paris 75005, France}
\affiliation{Université de Brest, CNRS, Ifremer, UMR6538 Geo-Ocean, 29280 Plouzané, France}
\affiliation{Lab-STICC -- UMR 6285 CNRS, ENSTA IP Paris, Brest, France}

\author{Axel Marmoret}
\affiliation{IMT Atlantique, Lab-STICC, UMR 6285 CNRS, Brest, France}

\author{Dorian Cazau}
\affiliation{Lab-STICC -- UMR 6285 CNRS, ENSTA IP Paris, Brest, France}

\author{Anatole Gros-Martial}
\affiliation{Centre d'Etudes Biologiques de Chizé (CEBC), UMR 7372, CNRS-La Rochelle Université, Villiers-en-Bois, France}

\author{Richard Dreo}
\affiliation{Université de Paris, Institut de physique du globe de Paris, CNRS, F-75005 Paris, France}
\affiliation{SAS Boksound, Kernevez, 20590 Rosnoen, France.}

\author{Maëlle Torterotot}
\affiliation{Lab-STICC -- UMR 6285 CNRS, ENSTA IP Paris, Brest, France}

\author{Sara Bazin}
\affiliation{Université de Brest, CNRS, Ifremer, UMR6538 Geo-Ocean, IUEM, 29280 Plouzané, France}

\author{Flore Samaran}
\affiliation{Lab-STICC -- UMR 6285 CNRS, ENSTA IP Paris, Brest, France}

\author{Jean-Yves Royer}
\affiliation{Université de Brest, CNRS, Ifremer, UMR6538 Geo-Ocean, 29280 Plouzané, France}

\email{pierre-yves.raumer@gmail.com}

\begin{abstract}
Passive hydroacoustic monitoring often generates large volumes of continuous recordings that are only partially exploited due to the cost of manual annotation. Supervised detection methods perform well but require large labeled datasets, seldom available for rare signals or understudied environments. This work proposes a self-supervised exploration pipeline to address this limitation in low-frequency settings.
A Masked AutoEncoder (MAE) is pre-trained on a reconstruction pretext task, then used to extract patch-level representations from spectrograms. Within each spectrogram, adjacent informative patches are aggregated into event-level embeddings, enabling the disentanglement of overlapping events. These embeddings are then clustered at the dataset scale using the dimension reduction algorithm UMAP and the clustering algorithm HDBSCAN to identify hydroacoustic patterns.
The pipeline was applied to a multi-year hydroacoustic dataset collected near Mayotte Island, Indian Ocean, containing marine mammal vocalizations, seismo-volcanic signals, and anthropogenic noise. The 317 clusters were manually mapped to 15 hydroacoustic classes or noise in less than one hour. The method was evaluated in two ways. Quantitatively, when used as a classifier, it achieved performance comparable to two existing detectors. Qualitatively, it recovered known seasonal patterns of marine mammal acoustic activity. It also identified patterns of previously unstudied signals, thereby demonstrating its practical value.

\end{abstract}
\maketitle
\nolinenumbers

\section{Introduction}

Passive hydroacoustic monitoring has become a key tool for observing the ocean at large spatial and temporal scales. Because low-frequency sound propagates efficiently underwater, hydrophone networks can record hydroacoustic activity over thousands of kilometers \citep[e.g.][]{ingale_hydroacoustic_2021}. Such datasets provide valuable information on marine mammals \citep[e.g.][]{torterotot_distribution_2020}, geophysical activity \citep[e.g.][]{raumer_automatic_2025}, and anthropogenic noise \citep[e.g.][]{haver_variable_2023}. Driven by these motivations, long-term hydrophone observatories have been deployed worldwide, and collected large volumes of continuous recordings.

Despite this abundance of data, only a small fraction of hydroacoustic recordings has been explored. Existing methodologies generally suffer from a trade-off between human effort and detection flexibility. On one hand, manual annotation is time-consuming, user-dependent, and scales poorly \citep[e.g.][]{raumer_open_2024, dubus_better_2023}. On the other hand, automatic detectors designed to assist these efforts either rely on large labeled datasets in the case of supervised training \citep[e.g.][]{dubus_citizen_2024, raumer_open_2024}, or avoid the training phase at the cost of specificity, requiring high prior knowledge to target restricted and structured signals \citep[e.g.][]{socheleau_automated_2015,torterotot_detection_2019, dreo_singing_2025}. Crucially, both automated paradigms confine exploration to predefined targets. This reliance on prior knowledge or annotated corpora limits the ability to rapidly explore datasets at scale and prevents the discovery of previously unknown hydroacoustic events.

In this work, we introduce a self-supervised exploration pipeline designed for large hydroacoustic datasets. The proposed method combines self-supervised representation learning with event-level clustering to identify recurring hydroacoustic patterns while requiring minimal manual annotation. Following the successful adaptation of transformers to vision learning tasks \citep{dosovitskiy_image_2020}, a Vision Transformer (ViT) is trained in a Masked AutoEncoder (MAE) framework \citep{he_masked_2021} on a large dataset. Because soundscapes are expected to vary across time and locations, the ViT pre-training is followed by an in-domain specialization training on a smaller target dataset. Within each spectrogram, adjacent patches are then aggregated into event-level embeddings, which are finally clustered at the dataset level.

We evaluate the proposed approach on the multi-year low-frequency hydroacoustic dataset MAHY collected near Mayotte Island in the Indian Ocean. The resulting clusters are used to construct a lightweight classification framework that requires only limited manual inspection. We propose this pipeline as a detector of marine mammal population-level acoustic presence and evaluate it as a classifier, showing that it achieves performance comparable to that of two existing detectors. In addition, the clusters enable the reconstruction of seasonal acoustic activity patterns for several known and unknown vocalizations and seismic signals, qualitatively illustrating the potential of the method as a tool for large-scale exploratory analysis of passive hydroacoustic monitoring data with minimal prior knowledge.

This work therefore provides a general methodology for exploring large hydroacoustic datasets with minimal annotation effort, supporting both the construction of approximate detectors and the discovery of recurring acoustic patterns. The main contributions of the paper are:
\begin{itemize}
    \item The adaptation of a self-supervised MAE framework to low-frequency hydroacoustic spectrograms for patch-level representation learning.
    \item An event-level disentanglement strategy that clusters adjacent patches within each spectrogram before dataset-scale clustering.
    \item A practical minimal-annotation exploration pipeline validated on a multi-year hydroacoustic dataset containing biological, geophysical, and anthropogenic signals.
\end{itemize}

\section{Related Works}

While various scientific fields study low-frequency hydroacoustic transient events, a common first step in their analysis is the construction of event catalogs. These catalogs may be built either manually or automatically. They may describe individual event occurrences (required, for instance, for source localization in seismology \citep[e.g.][]{tsang-hin-sun_seismicity_2016, ingale_hydroacoustic_2021}), or periods of binary presence/absence. The latter is useful for assessing marine mammal acoustic presence, which in turn allows for the study of the spatio-temporal distribution of their populations \citep[e.g.][]{torterotot_distribution_2020}.

To automate event detection, many traditional approaches leverage prior knowledge of events to design task-specific detectors, such as energy detectors (e.g. STA/LTA), shape-based detectors \citep[e.g.][]{samaran_seasonal_2013,leroy_reliability_2018} or periodicity-based detectors \citep{dreo_singing_2025}. In recent decades, data-driven methods have proven accurate, in particular supervised detectors based on Convolutional Neural Networks (CNN) \citep[e.g.][]{dubus_citizen_2024, raumer_open_2024, schall_deep_2024}.

Recently, self-supervised learning (SSL) has emerged as a powerful alternative. Rather than relying on manual labeling, SSL leverages large amounts of unlabeled data to learn representations that reflect the underlying structure of the audio signal. These representations are learned during a pre-training phase via pretext tasks, such as reconstructing an artificially corrupted input or contrasting similar and dissimilar samples. The model thus derives compact, generic, and structurally rich representations of the data, known as embeddings, that can then be applied to downstream tasks such as classification, clustering, or anomaly detection. When a pre-trained model is subsequently fine-tuned on a small amount of labeled data for classification, the procedure falls under the semi-supervised learning paradigm. In bioacoustics, such approaches, relying on reconstruction or contrastive objectives, have already been applied to architectures including CNNs, Transformers, and State Space Models (SSMs) \citep[e.g.][]{baker_audiosam_2025,rauch_can_2025,tang_state_2025}. Beyond classification, the learned embeddings have also been used directly for anomaly detection \citep[e.g.][]{bermant_bioacoustic_2022} and for clustering \citep[e.g., in seismology;][]{rimpot_selfsupervised_2025}. This work builds upon these approaches but, rather than clustering whole-window embeddings, it extracts event-level representations within each spectrogram before dataset-level clustering, partly addressing the problem posed by the frequent overlap of sources in low-frequency hydroacoustic data.

\section{Masked Autoencoders}

\begin{figure*}[!h]
\centering
\includegraphics[width=14cm]{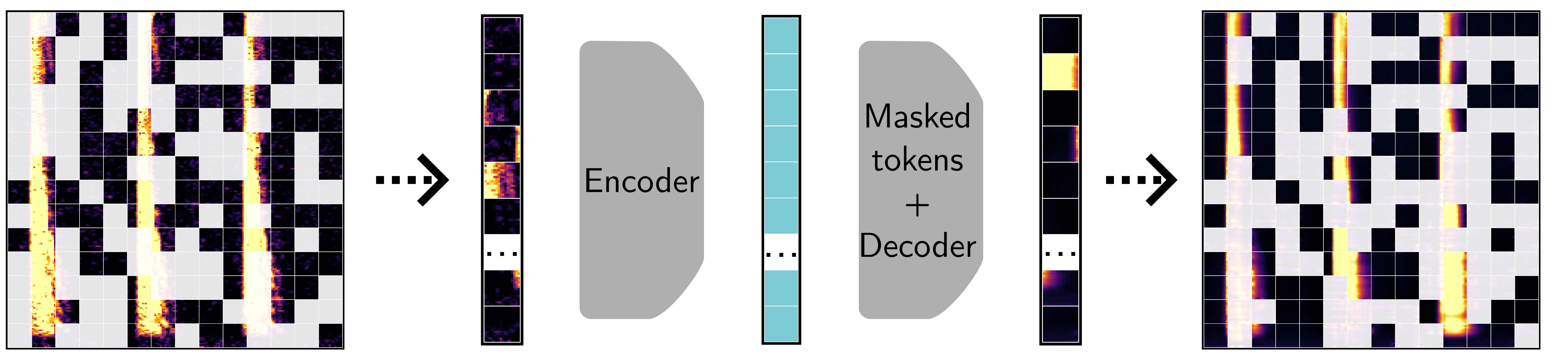}
\caption{Schematic view of the architecture used in this work. The input spectrogram is partly masked, then (arrows) reshaped to be fed to the ViT. An encoder then embeds each patch to a 256-dimensional vector (blue rectangle), after which masked tokens are injected to specify the position of the patches that were masked in the input. Lastly, the decoder reconstructs missing patches and (arrow) a reshaping operation allows the result to be visualized.}
\label{fig_MAE-schema}
\end{figure*}

The Masked AutoEncoder (MAE) is a reconstruction-based architecture introduced by \citet{he_masked_2021}. It learns to reconstruct an image from a randomly masked version: portions of the input are hidden, and the model is trained to predict the missing parts from the visible ones. In the field of audio, the input image can be a spectrogram. To reconstruct it, the model must learn the inherent structure of the acoustic environment, such as the typical time-frequency signature of specific animal calls, which makes the approach well suited to working directly on spectrograms \citep[e.g.][]{baker_audiosam_2025, rauch_can_2025}.

In this work, as in the original paper, the encoder and decoder are Vision Transformers \citep[ViT-MAE,][]{dosovitskiy_image_2020}, forming an asymmetric encoder–decoder. The input spectrogram is divided into patches, a fraction of which is masked. The encoder embeds each unmasked patch into a dimensional vector; masked tokens are then injected to specify the positions of the masked patches, and the decoder reconstructs them. The model is trained using a mean-squared error (MSE) loss computed only over the masked patches:
\begin{equation}
L = \frac{1}{N_c} \sum_{i=1}^{N_c} \frac{1}{M^2} \sum_{j=1}^M\sum_{k=1}^M (y_{i_{jk}}-\widehat{y_{i_{jk}}})^2
\label{eq_MSE}
\end{equation}
where $L$ is the MSE loss, $N_c$ the number of masked patches, $M$ the side of a patch, $y_{i_{jk}}$ the grayscale value of the pixel at row $j$ and column $k$ of patch $i$, and $\widehat{y_{i_{jk}}}$ its predicted counterpart. Figure \ref{fig_MAE-schema} shows a schematic view of the MAE architecture used, with an actual spectrogram example.

\section{Data}
\begin{figure*}[!h]
\centering
\includegraphics{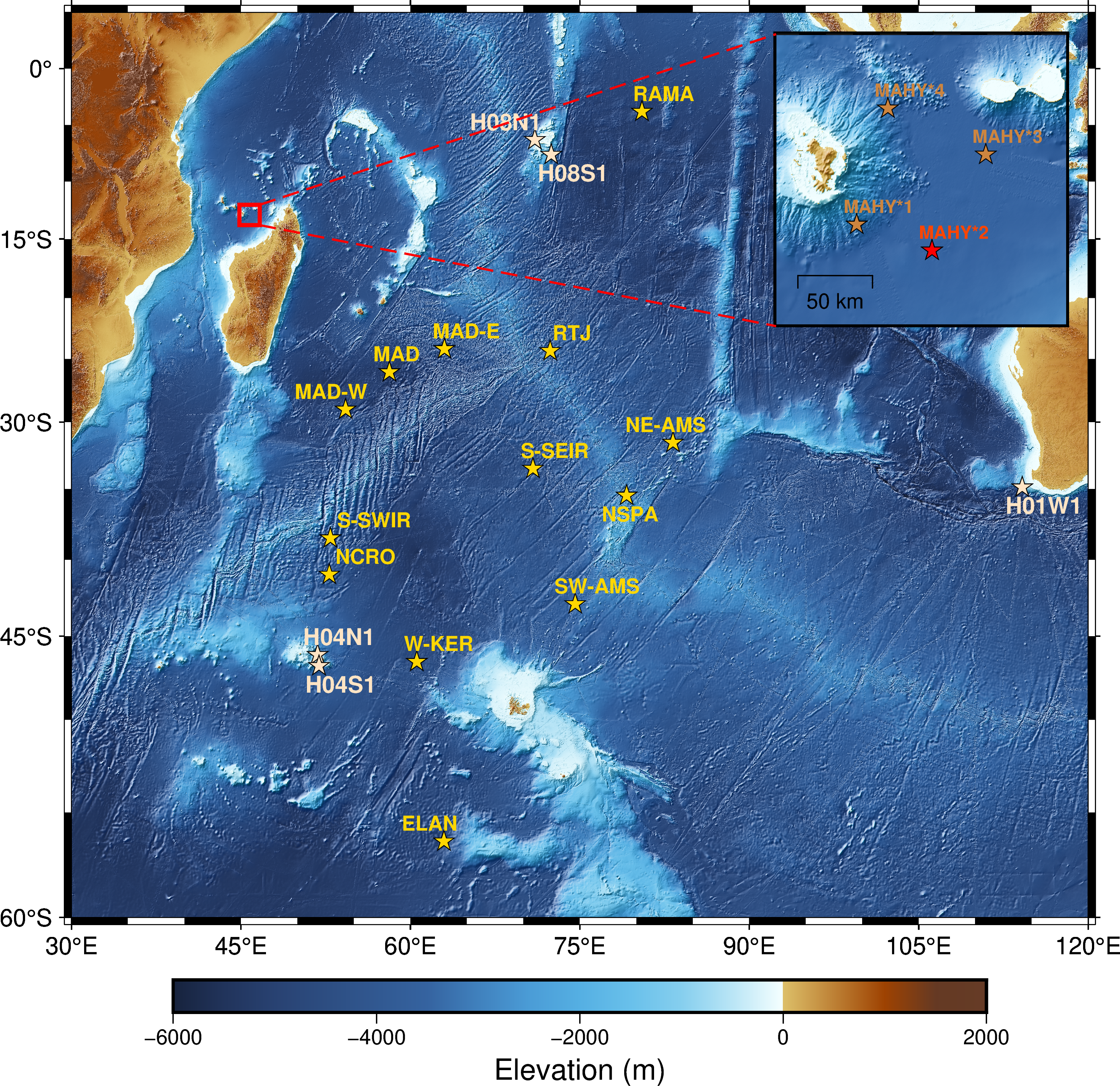}
\caption{Map of the southern Indian Ocean showing the locations of the hydrophones used in this work. Yellow stars: hydrophones from OHASISBIO. Beige stars: cabled hydrophones from IMS (CTBTO). Brown stars: MAHY hydrophones included in the pre-training dataset. Red star: MAHY*2, the hydrophone targeted by the experiment presented in this paper.}
\label{fig_hydro_map}
\end{figure*}

The model was pre-trained on a comprehensive dataset compiled from several hydrophone networks in the Indian Ocean, spanning from 2009 to 2024. This corpus includes recordings from two autonomous observatories, OHASISBIO in the southern Indian Ocean \citep{royer_oha-sis-bio_2009} and MAHY near Mayotte Island \citep{bazin_initial_2023}, and cabled hydrophones operated by the Comprehensive Nuclear-Test-Ban Treaty Organization (CTBTO). All instruments were moored within the SOFAR channel, recording at native sampling rates of 240 or 250\,Hz. To ensure acoustic consistency across datasets, all time series were systematically resampled to a uniform 240\,Hz. This large-scale, multi-year collection encompasses highly diverse locations, seasons, and ambient environments, providing a rich variety of hydroacoustic signals for self-supervised learning. Figure \ref{fig_hydro_map} shows the distribution of the stations that were used in this work.

After the initial SSL phase, the in-domain specialization and the clustering methodology that will be presented in this work were applied to a single hydrophone from the MAHY network, hereafter referred to as MAHY*2. This hydrophone was chosen because its data had already been partly annotated for marine mammal vocalizations \citep{dreo_singing_2025} and analyzed regarding seismo-volcanic signals from the Fani Maoré volcano \citep{lavayssiere_hydroacoustic_2024, raumer_automatic_2025, bazin2025}. In terms of seismo-volcanic activity, the network recorded signals comprising direct seismic phases and hydroacoustic waves generated by interactions between lava and seawater \citep{bazin2025}. In terms of bioacoustic activity, \citet{dreo_singing_2025} have reported vocalizations from several baleen whale species, including Antarctic blue whales (\textit{Balaenoptera musculus intermedia}; ABW), South Western Indian Ocean pygmy blue whales (\textit{Balaenoptera musculus brevicauda}; SWIO-PBW), Antarctic minke whales (\textit{Balaenoptera bonaerensis}; AMW), and fin whales (\textit{Balaenoptera physalus}; FW). The possible presence of Omura’s Whale (\textit{Balaenoptera omurai}; OW), a species only recently described and still rarely identified in the field \citep[e.g.][]{cerchio_global_2019}, has also been suggested. Lastly, while they were not the primary target of annotation, many anthropogenic signals such as ship noise, airgun or quarry shots were observed in the recordings \citep[e.g.][]{raumer_automatic_2025}.

The coexistence of known geophysical, biological, and anthropogenic signals in these long-term continuous recordings makes the MAHY dataset particularly well-suited for evaluating unsupervised exploration methods. The diversity of recurring hydroacoustic events and their seasonal variability provide a realistic case study for assessing whether clustering-based approaches can identify meaningful signal classes and estimate their temporal distribution with minimal manual annotation.

\section{Methodology}
\begin{figure*}[!h]
\centering
\includegraphics{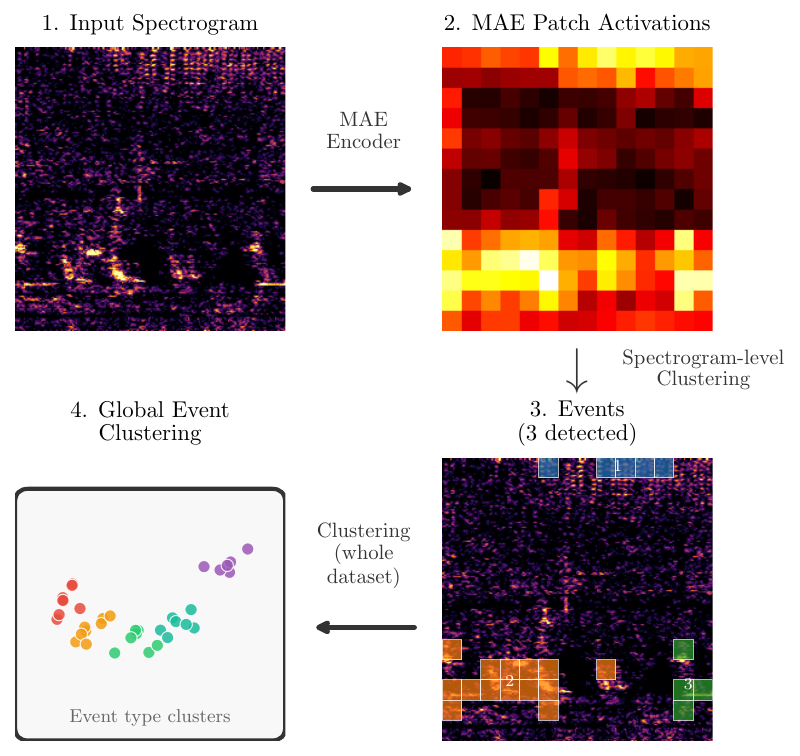}
\caption{Overview of the proposed workflow. (1) A spectrogram containing both a 100-120 Hz AMW call and  15-40 Hz SWIO-PBW calls. The MAE produces an embedding vector for each patch, whose L2 norm is shown in (2). (3) Spectrogram-level clustering enables the separation of distinct hydroacoustic events within the time window. (4) Event-level embedding vectors are then clustered at the dataset level. In the example, events 2 and 3 from step (3) both account for SWIO-PBW calls and are expected to lie in the same cluster $\alpha$, whereas event 1 represents a different signal class and is expected to lie in another cluster $\gamma$.}
\label{fig_method-overview}
\end{figure*}

The proposed methodology comprises four steps. First, a representation of the data is learned using a Masked AutoEncoder (MAE). Secondly, event-level features are extracted by clustering patch embeddings within individual spectrograms. Thirdly, the extracted event embeddings are clustered at the dataset level to identify recurring hydroacoustic patterns. Lastly, visually similar clusters are grouped together following a brief manual annotation phase. Figure \ref{fig_method-overview} gives a schematic overview of the first three points.

\subsection{Masked-Autoencoder Training}

Targeting low-frequency soundscapes, the model handles 128\,s audio data sampled at 240\,Hz. With a STFT window size of 480 samples and an overlap of 75\%, temporal and spectral resolutions of 0.5\,s and 0.5\,Hz are obtained. Spectrogram whitening is then performed frequency-wise by dividing each time-frequency bin by the local temporal mean of its corresponding frequency, estimated over a 500\,s window centered on the considered segment, thereby accounting for ambient noise variations. The resulting values are clipped between 1 and 30, corresponding to 0 and $\sim$15 dB. This removes values smaller than the background noise while ensuring all high-energy values are kept. Lastly, the resulting images are rescaled to 224x224 and divided into 196 16x16 patches, of which 98 (50\%) are masked.

A model architecture close to the one of the original paper was used: the encoder is a ViT \citep{dosovitskiy_image_2020} with 16 layers, 16 attention heads, and an embedding dimension of 256. The decoder is a ViT with six layers and 16 attention heads. Overall, a relatively lightweight architecture was selected to limit computational requirements and ease the reproducibility on modest devices. The ViT was adapted to spectrograms by allowing for single-channel grayscale input, whereas the base model used 3-channels RGB inputs.
Given the patch-wise nature of the reconstruction pretext task, the model is expected to capture meaningful local features, particularly from recurrent stereotyped marine mammal vocalizations. Indeed, because these calls exhibit nearly constant acoustic signatures, they generate highly predictable and repeating patterns on spectrograms, that align well with the model's learning objective.

\subsection{Spectrogram-Level Event Extraction}
Applying the MAE encoder to a spectrogram produces a grid of 14×14 embedding vectors of dimension 256, each corresponding to a spectrogram patch. This time-frequency-wise embedding structure is exploited to detect individual transient hydroacoustic events and assign a representative embedding to each of them.

This is performed in two steps: first, ``empty'' patches are discarded. Then, the remaining patches are clustered according to their neighborhood.

\subsubsection{Patch Discarding}

For each patch $i$ of the spectrogram, we define an energy $E_i$ and an activation $A_i$ as follows:

\begin{equation}
E_i=\frac{1}{M^2} \sqrt{\sum_{j=1}^M\sum_{k=1}^M y_{i_{jk}}^2}
\label{eq_E}
\end{equation}

\begin{equation}
A_i=\frac{1}{H} \sqrt{\sum_{j=1}^H h_{i_j}^2}
\label{eq_A}
\end{equation}

With $h_{i_j}$ being the $j^{\text{th}}$ component of the embedding vector associated to patch $i$.

A patch $i$ is defined as ``empty'' if at least one of two conditions is fulfilled:
\begin{itemize}
    \item $E_i < \mu_E - \sigma_E$, where $\mu_E$ and $\sigma_E$ are the spectrogram-wise average and standard deviation of energies.
    \item $A_i < \mu_A + \sigma_A$, where $\mu_A$ and $\sigma_A$ are the spectrogram-wise average and standard deviation of activations.
\end{itemize}

This methodology serves a double purpose: discarding regions dominated by background noise, and regions that poorly contribute to the MAE reconstruction process. Indeed, higher embedding norms were empirically observed to correspond to structurally informative patches of the spectrogram. The thresholds were chosen empirically as a compromise between missing events and selecting noise. 

\subsubsection{Spectrogram-Level Patch Clustering}

Once the removal process is complete, only a few regions should remain. Our goal is to merge the remaining adjacent regions, which we will refer to as “events” from here on. To do this, we use the following clustering algorithm: initially, each remaining region is considered an independent event. Next, any two events $K$ and $L$ are merged if their minimum Chebyshev distance is less than or equal to three, that is if:

\begin{equation}
\min_{(x,y)\in K, (x',y')\in L} \max(|x-x'|,|y-y'|) \le 3.
\label{eq_chebychev}
\end{equation}
Here, $(x,y)\in K$ is the position of patches belonging to event $K$, i.e., row $x$ and column $y$. In practice, this condition merges any two events that are separated by a gap of two patches or fewer, counting a patch diagonal as a single step.

This merging operation is iteratively repeated until no further merges are possible. Figure \ref{fig_method-overview} gives an example of a spectrogram-level patch clustering process resulting in three different clusters.
Although this method causes adjacent real events in the spectrograms to merge into a single group, it should still be able to distinguish non-adjacent events.

Finally, for each event, the mean embedding vector is computed. The patch with the embedding vector being the closest to this average, in terms of cosine distance, is then selected as the representative embedding and stored along with the event size and mean time-frequency coordinates. This strategy avoids using artificial centroid vectors that do not correspond to actual observations and ensures that all stored embeddings correspond to real patches.

\subsection{Dataset-Level Clustering}
After extracting all event-level features over the entire target dataset, representative embeddings are pooled into a single dataset for clustering. Two methods are compared:
\begin{itemize}
    \item \textit{MAE-K-Means}: Following \cite{rimpot_selfsupervised_2025}, a first application of K-Means \citep{lloyd_least_1982} on the embeddings is applied to obtain 2,000 clusters. A standard agglomerative hierarchical clustering \citep{johnson_hierarchical_1967} is then used to merge neighboring clusters until the desired number of clusters is reached. This second step enables to reduce the number of clusters to a manageable size, more specifically to ensure a human inspection is feasible.
    \item \textit{MAE-UMAP}: A UMAP \citep[Uniform Manifold Approximation and Projection,][]{mcinnes_umap_2018} model first reduces the dimensionality of the embedding vectors from 256 to 16. HDBSCAN \citep[Uniform Manifold Approximation and Projection,][]{hutchison_density-based_2013} is then applied to the projected features.
\end{itemize}

We chose \textit{MAE-K-Means} as the reference model to produce the figures of this paper and to attempt ablation studies.

\subsection{Cluster-to-Class Mapping}

Dataset-level clusters are not expected to provide a 1-to-1 mapping to actual signal classes. For example, airgun shots may vary in period and shot duration, likely resulting in several clusters. To solve this issue, an initial lightweight annotation step is required to associate clusters with semantic classes. This cluster-to-class mapping is done in this way: for each cluster, eight spectrograms are randomly selected among the 10\% closest samples to the cluster centroid, ensuring that the inspected examples are representative of the cluster. These samples are visually examined by an annotator. Because the dataset-level clustering groups events rather than full spectrograms, the time-frequency position of each event in the spectrogram is known. This position is indicated to the annotator by two dashed lines to facilitate interpretation. Clusters exhibiting visually similar acoustic patterns are then grouped into broader hydroacoustic classes, named after their shape or source in the case where it is known by the annotator.

\section{Evaluation Protocol}

We evaluated the proposed clustering-based pipeline using a combination of qualitative and quantitative analyses on the MAHY*2 dataset.

\subsection{Classification Framework}
To enable quantitative comparison with existing detection methods, the clustering results were interpreted within a classification framework. Ground-truth labels were obtained from two independent sources: (i) the weakly annotated marine mammal vocalization dataset of \citet{dreo_singing_2025}, and (ii) the automatic seismic P-phase catalog of \citet{raumer_automatic_2025}. Both provide hourly binary presence/absence labels.
Since the proposed pipeline operates on 128-second spectrogram windows, predictions were aggregated on an hourly basis.

\subsection{Evaluation Metrics} \label{sec:scoring}
To transform dataset-level clusters into detectors, we introduce an hourly density score. Let a semantic class $C$ be associated with one or more clusters, and consider a given hour $h$. We denote by $n_C(h)$ the number of spectrograms within $h$ that are assigned to one of the clusters mapped to the class, and by $n(h)$ the total number of spectrograms recorded during $h$. We define the normalized hourly score as:
\begin{equation}
s(h) = \frac{n_c(h)}{n(h)}
\label{eq_thresh}
\end{equation}
This score reflects how strongly the class is expressed during the hour: it ranges from zero, when no spectrogram of the hour belongs to the class, to one, when all of them do, and may exceed one if more than one event is present per spectrogram on average.

By varying a threshold $\tau$ on $s(h)$, one obtains a family of binary hourly classifiers yielding for each hour the boolean $s(h)>\tau$.
This enables the construction of Receiver Operating Characteristic (ROC) curves by computing the True Positive Rate (TPR) and False Positive Rate (FPR) over the range of threshold values. For each threshold, we computed True Positive (TP), False Positive (FP), True Negative (TN), and False Negative (FN) counts at the hourly scale. From these, we derived standard metrics including TPR, FPR, precision, recall, and the F1-score, defined as $\frac{2 \cdot \text{precision} \cdot \text{recall}}{\text{precision} + \text{recall}}$.

\subsection{Baseline Comparisons and Ablations}

The proposed method was compared against two alternative approaches:

\begin{itemize}
\item A periodicity-based detector developed by \citet{dreo_singing_2025}, specifically designed for baleen whale stereotyped vocalizations.
\item A supervised object-detection model based on YOLO, developed as a baseline of the BioDCASE challenge \citep[code available at][]{marinebioCASE2025}.
\end{itemize}

In addition to the two event-level clustering strategies, variations of our method ablating the event-level extraction stage were evaluated. Instead, a single embedding was computed per spectrogram using different aggregation strategies at the spectrogram level:
\begin{itemize}
    \item Average of embedding vectors.
    \item Embedding vector of maximal activation patch with the activation defined by Equation \ref{eq_A}.
    \item Embedding vector of the patch with the maximum absolute deviation from the spectrogram embedding centroid.
\end{itemize}

Moreover, to investigate UMAP relevance for \textit{MAE-UMAP}, an equivalent version replacing UMAP with PCA was attempted, keeping the same HDBSCAN input dimension and parameters.

Lastly, an additional ablation of the pre-training step has been evaluated. Instead, the MAE model is directly trained on the MAHY*2 dataset.

All those alternative methods were followed by clustering and classification using the same procedure as \textit{MAE-K-MEANS}.

This comparison framework allows us to evaluate both the value of the event-level representation and that of the pre-training, within a supervised detection framework.

\subsection{Qualitative Evaluation}

In addition to the quantitative assessment above, we analyzed the temporal distribution of detections for each class. Because many hydroacoustic signals, particularly baleen whale vocalizations, exhibit strong seasonal patterns, meaningful clustering is expected to produce structured temporal distributions. Moreover, the number of earthquakes in a seismic sequence is expected to decrease over time \citep{raumer_automatic_2025}. The agreement between observed and expected patterns provides a qualitative validation of the method. However, the lack of prior knowledge causes no expectation regarding ship noise, quarry or airgun shot patterns.

\section{Results}
\subsection{MAE Training}

For pre-training, a subset of 9,887,277 spectrograms was randomly extracted from 34 different geographical sites, in order to reduce the pre-training time while maintaining diversity of examples. Generated using a 50\% overlap, this subset represents 12.20\% of the total available data. The exhaustive set of 1,137,732 spectrograms from MAHY*2 was then used for in-domain specialization, with a similar overlap. The process required two weeks on a single GPU on a cluster for pre-training and one day using an NVIDIA RTX PRO 1000 Blackwell on a laptop for both in-domain specialization and pre-training ablation learning phase. 80\% of the spectrograms were used for training, and the remaining 20\% for validation with the same MSE loss, both for pre-training and in-domain specialization when applicable. Training was stopped by early stopping, with a patience of two epochs, leading to 12 and 9 epochs for pre-training and in-domain specialization, respectively, and to 31 epochs with the pre-training ablated.

\begin{figure*}[!h]
\centering
\includegraphics{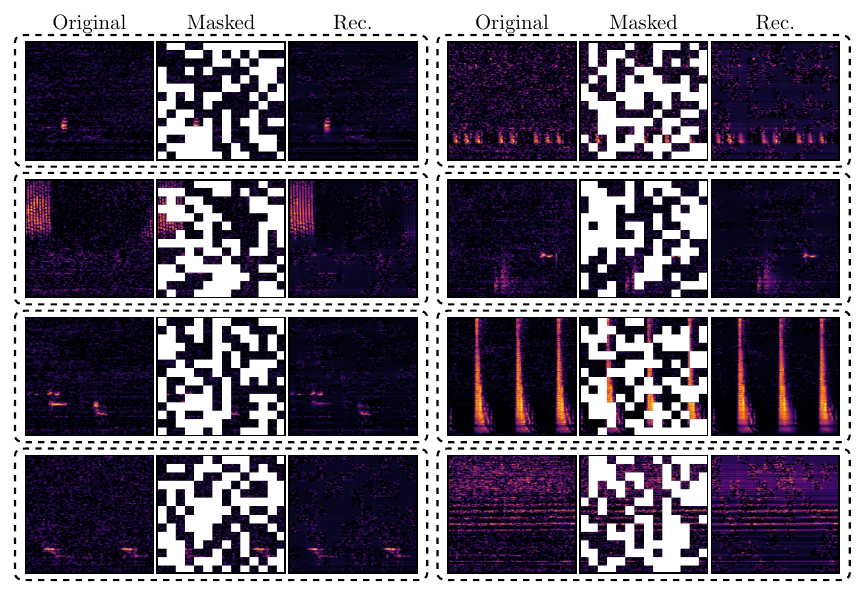}
\caption{Examples of MAHY*2 normalized spectrograms reconstructions by the MAE. Original stands for the base spectrogram, Masked shows the 50\%-masked counterpart given as input to the model, and Rec. is the unmasked input added to the masked patch reconstruction. Spectrograms are normalized per example.}
\label{fig_MAE-perf}
\end{figure*}

Figure \ref{fig_MAE-perf} shows examples of spectrograms given as input to the MAE model, together with a 50\,\% masked counterpart and the output reconstruction from the MAE.

\subsection{Dataset-Level Clustering}
To ensure computational tractability, a subset of 300,000 randomly selected embedding vectors was first used to fit the UMAP and HDBSCAN or K-Means and agglomerative clustering, after which the remaining samples were assigned to existing clusters. Hyperparameters were tuned to obtain a number of clusters on the order of 300 for \textit{MAE-UMAP}. This led to fixing the minimum cluster size of 50 and a minimum neighborhood size of 20. This choice is a trade-off between the presence of a few large clusters and the scattering of the dataset into many small clusters, likely more fine-grained but requiring a long annotation time. A total of 317 clusters were obtained, containing between 187 and 21,222 events, with a median size of 1,304. One hour of visual inspection led the annotator to identify 13 classes. 

To enable comparison, the maximum merging distance of 0.09602 was used to limit the agglomerative clustering of \textit{MAE-K-Means}, to obtain the exact same number of clusters as \textit{MAE-UMAP}. A total of 317 clusters were thus obtained, containing between 237 and 300,594 events, with a median size of 2,328. One hour of visual inspection led the annotator to identify 15 classes, entirely including the 13 classes from \textit{MAE-UMAP}. To describe the classes and provide temporal distributions of classes, the results of \textit{MAE-K-Means} will be used.

Table \ref{table_classes} provides a list of the 15 identified classes, along with the number of clusters that were mapped to these classes and the total number of events belonging to them using \textit{MAE-K-Means}. A similar table is given for \textit{MAE-UMAP} in supplementary materials. An example is given for each class in Figure \ref{fig_histograms}. Among the classes, five were of unknown origin, while the other ten were attributed to known sources. Two unknown classes, ``Und. 42 Hz'' and ``LF 8 sec pulse'', were already observed close to Mayotte, while not extensively discussed \citep{dreo_singing_2025}.

\begin{table*}[ht]
\caption{Classes selected by the annotator while browsing the clusters yielded by \textit{MAE-K-Means}. 75\% f (Hz) and 75\% d (s) show the 75\% quantiles of frequencies and durations, using rounded multiples of the patch time and frequency resolution, 9.1\,s and 8.6\,Hz respectively. Note the 87 missing clusters were simply discarded because associated to unstructured noise, and contained 2,629,623 events out of the 3,485,990.}
\label{table_classes}
\begin{ruledtabular}
\renewcommand{\arraystretch}{0.85}
\setlength{\tabcolsep}{3.5pt}
\begin{tabular}{l l cc cc}
Class & Description & 75\% f (Hz) & 75\% d (s) & N clusters & N events \\
\hline
ABW\_Z         & Antarctic Blue Whale Z-calls & 17-26 & 9-36 & 16 & 39993 \\
AMW            & Antarctic minke whale calls & 94-111 & 18-73 & 20 & 93289 \\
Airgun         & Seismic airgun survey & 0-111 & 9-18 & 35 & 33323 \\
BW\_D          & Blue Whale D-calls & 26-103 & 9-27 & 8 & 18225 \\
EQ\_P          & Seismic P-phases & 0-34 & 18-54 & 13 & 43013 \\
EQ\_LF          & Very low frequency seismic phase & 0-26 & 45-100 & 2 & 1885 \\
FW\_20         & Fin whale 20\,Hz pulses & 9-26 & 18-109 & 7 & 33942 \\
HF tones       & Unknown signal of high frequency & 94-103 & 9-45 & 3 & 6665 \\
Impulsive      & Unknown impulsive signal & 17-111 & 9-27 & 4 & 6512 \\
LF 8 sec pulse & Unknown signal of 8\,s & 51-77 & 9-45 & 2 & 23059 \\
OW             & Omura's whale calls & 26-34 & 9-18 & 9 & 10068 \\
SWIO-PBW       & Madagascar pygmy blue whale calls & 9-26 & 9-45 & 55 & 331975 \\
Ship           & Ship noise & 34-94 & 9-54 & 39 & 184199 \\
Und. 20 Hz     & Unknown signal at 20\,Hz & 9-34 & 18-36 & 1 & 1530 \\
Und. 42 Hz     & Unknown signal at 42\,Hz & 34-43 & 9-27 & 16 & 28689 \\
\end{tabular}
\end{ruledtabular}
\end{table*}

\subsection{Temporal Distribution of Classes}

\begin{figure*}[!h]
\centering
\includegraphics{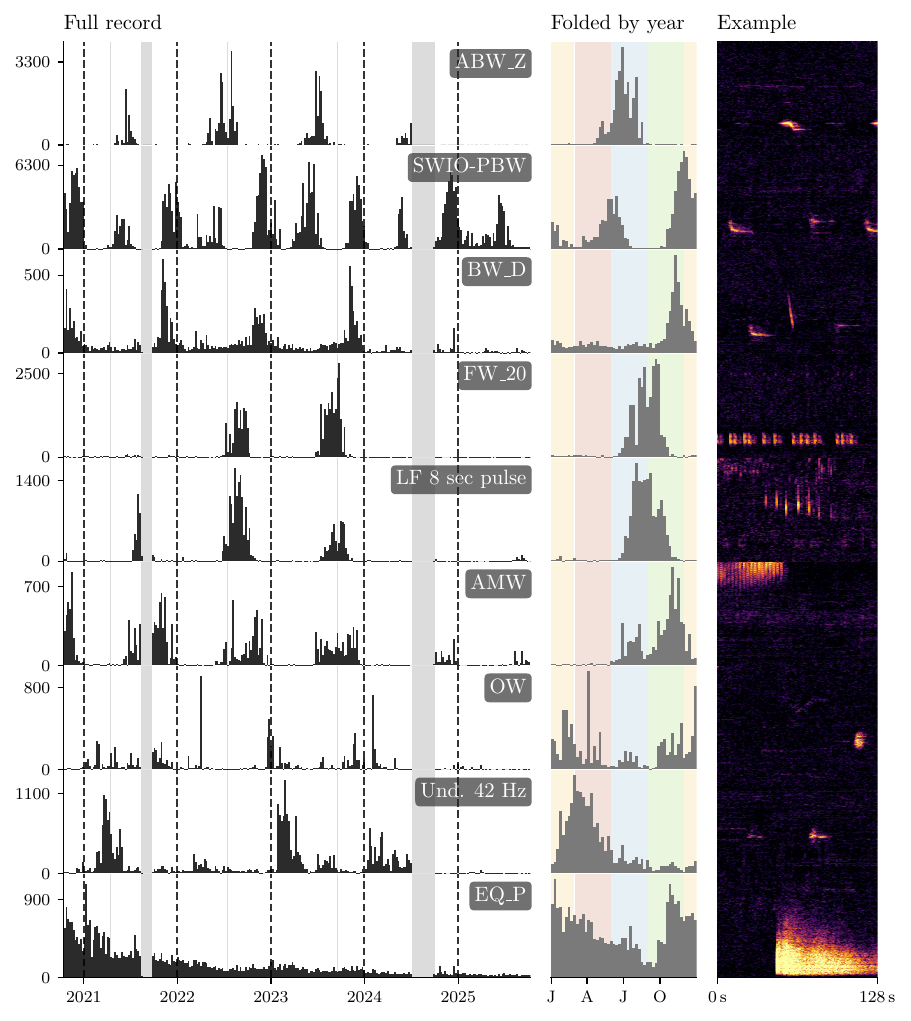}
\caption{Time distribution of detected events with an example spectrogram, using the results of \textit{MAE-K-Means}. Each histogram bin shows the number of events detected per week. The histogram displays the full temporal distribution (black) and its yearly normalized average for seasonality (gray). Background colors indicate data gaps (gray) and seasons in the Southern Hemisphere (yellow: summer, red: autumn, blue: winter, green: spring). The spectrograms are normalized and span 0--120\,Hz.}
\label{fig_histograms}
\end{figure*}

Figure \ref{fig_histograms} gives examples of temporal distributions of events from nine classes. The distributions of the other classes, not shown in this figure because they were deemed less useful for qualitative assessment, are presented in the supplementary materials.

The histograms of marine mammal-related classes show clear seasonal structure except for ``OW''. Notably, the SWIO-PBW shows two annual acoustic presence peaks. Although the ``FW\_20'' class exhibits a seasonal pattern, it is absent in 2021, 2024 and 2025. The 2024 absence coincides with a data gap. The ``OW'' class of biological origin does not show any clear seasonal pattern. Lastly, ``EQ\_P'' decreases during the recording period.

\subsection{Classification Performance}

\begin{figure*}[!h]
\centering
\includegraphics{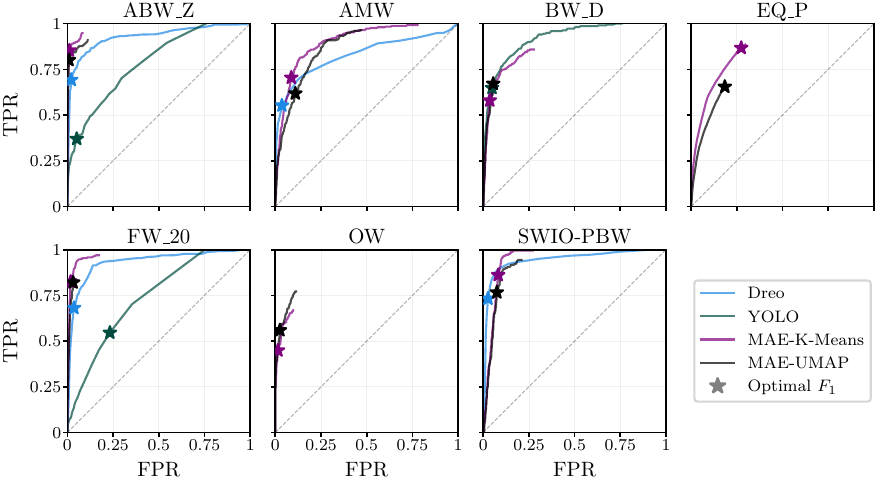}
\caption{Receiver Operating Characteristic curves of the presented MAE-based methods, the BioDCASE YOLO baseline, and the detector of \citet{dreo_singing_2025} for the seven classes for which ground truth was available. \textit{FPR}: False Positive Rate. \textit{TPR}: True Positive Rate. The stars represent the point of optimal F1 score (see Table \ref{table_eval}). The dotted line corresponds to the score a random classifier would have. The ground truth was obtained from the annotation campaign of  \citet{dreo_singing_2025} for whale calls, and from the automatic catalog of \citet{raumer_automatic_2025} for seismic P-phases.}
\label{fig_ROC}
\end{figure*}

Figure \ref{fig_ROC} shows the ROC curves of the proposed MAE-based methods, together with those of the baseline detectors, \citet{dreo_singing_2025} and the BioDCASE YOLO model, for the seven classes where annotations were available. As described in Section \ref{sec:scoring}, the curves of the proposed method do not span the full FPR range, because its decision region is bounded by the cluster support. The star marks the threshold maximizing F1. For clarity, only the main \textit{MAE-UMAP} and \textit{MAE-K-Means} methods are shown; the ablated variants are reported in Table \ref{table_eval} only. The performance of other MAE-based methods, with spectrogram-level clustering ablated, is lower than that of the proposed method. In particular, averaged embeddings yield the poorest performance. The performance of the PCA+HDBSCAN pipeline was particularly poor compared to other techniques, in particular, the cluster to class mapping only enabled to identify one of the evaluated classes. The results are given in supplementary materials.

\begin{table*}[ht]
  \caption{Evaluation metrics for each class.
    P: number of positive samples, N: number of negative samples.
    F1: optimal hourly F1-score (\%). It is to be noted that this is a measure of the achievable performance of a model.
    Clustering variants: K-Means (Agglomerative clustering on K-Means, the \textit{MAE-K-Means} model), UMAP (HDBSCAN after UMAP, the \textit{MAE-UMAP} model).
    Representation variants of \textit{MAE-UMAP}: avg (mean embedding), max1 (max norm), max2 (farthest from centroid). Pre-training variant: no pre-train (direct training on MAHY*2)}
  \label{table_eval}
  \centering
  \renewcommand{\arraystretch}{1.3}
  \setlength{\tabcolsep}{3pt}
  \begin{tabular}{l rr cc cc ccc c}
    \hline
    & \multicolumn{2}{c}{Samples}
    & \multicolumn{2}{c}{Baselines ($F_1$ \%)}
    & \multicolumn{6}{c}{MAE Method Variants ($F_1$ \%)} \\
    \cline{2-3} \cline{4-5} \cline{6-11}
    & & & & & \multicolumn{2}{c}{Clustering} & \multicolumn{3}{c}{Representation} & \multicolumn{1}{c}{Pre-training} \\
    \cline{6-7} \cline{8-10} \cline{11-11}
    Class & P & N & Dreo & YOLO & K-Means & UMAP & avg & max1 & max2 & no pre-train \\
    \hline
    ABW\_Z & 211 & 2\,172 & 72.57 & 39.00 & \textbf{86.75} & 84.71 & --- & --- & 85.64 & 83.46 \\
    AMW & 254 & 2\,129 & \textbf{59.55} & --- & 57.84 & 48.76 & 42.41 & 1.78 & 47.44 & 51.15 \\
    BW\_D & 219 & 2\,164 & --- & \textbf{61.08} & 60.05 & 60.74 & --- & --- & 51.12 & 48.78 \\
    EQ\_P & 12\,975 & 19\,575 & --- & --- & 76.06 & 67.76 & 41.28 & 0.31 & \textbf{83.84} & 75.39 \\
    FW\_20 & 262 & 2\,121 & 69.07 & 31.85 & \textbf{83.72} & 79.04 & 9.42 & 70.20 & 79.48 & 81.25 \\
    OW & 118 & 2\,265 & --- & --- & 51.96 & \textbf{54.77} & --- & --- & 46.86 & 46.81 \\
    SWIO-PBW & 180 & 2\,203 & \textbf{71.44} & --- & 60.19 & 57.26 & 30.98 & 16.70 & 59.33 & 61.86 \\
    \hline
  \end{tabular}
\end{table*}

\section{Discussion}
The proposed method was evaluated quantitatively by placing it within a classification framework. When used as a classifier, the method reaches performance levels equivalent to or better than those of the baseline detectors except for one class. It should be noted, however, that due to its very nature, it does not allow for adjustments to the trade-off between false positives and false negatives, which makes the generated ROC curves incomplete (Figure \ref{fig_ROC}). The results summarized in Table \ref{table_eval} further indicate that this approach is applicable to various categories of signals, including marine mammal vocalizations and seismic waves. Although the $F1$ scores obtained remain moderate and precision is limited for several classes, these values must be interpreted in light of the fundamentally unsupervised nature of the pipeline, which required only minimal annotation time. The objective of the proposed method is to provide an exploration tool accurate enough to approximate a classification process with minimal human intervention. In this respect, the quantitative evaluation demonstrates that it is possible to achieve significant detection capabilities with minimsl annotation effort.

In the MAHY*2 case study, approximately one hour was sufficient for an annotator to browse the clusters and assign them to broader hydroacoustic classes. This brief inspection stage provides access to a large number of automatically extracted events and allows you to quickly build time series of presence–absence data for various types of signals.

The resulting temporal distributions are highly consistent with the literature and independent catalogs. For instance, the recovered baleen whale seasonalities,including the 2021 and 2024 absence of ``FW\_20'', match previous observations by \citet{dreo_singing_2025}, while the decrease in ``EQ\_P'' aligns with expected seismic sequences \citep{raumer_automatic_2025}. Other unexpected patterns, such as the bi-annual presence of SWIO-PBW or the joint occurrence of ``FW\_20'' and the ``LF 8 sec pulse'', open up new avenues for biological interpretation. This agreement suggests that, even when per-window classification accuracy remains imperfect, aggregated cluster occurrences provide reliable estimates of seasonal hydroacoustic activity; capturing broad temporal structure may indeed be more critical than achieving optimal event-level precision.
Beyond validating known patterns, a key motivation of this unsupervised pipeline was the discovery of unknown signals. Since this pipeline does not rely on prior knowledge of the signals present in the dataset, it also makes it possible to detect hydroacoustic events that were previously unidentified or poorly documented. In the MAHY*2 example, five unknown classes were identified. While the classes ``LF 8 sec pulse'' and ``Und. 42 Hz'' were identified in previous annotation campaigns, they had not been characterized in detail, and, to our knowledge, do not appear in the literature. The clustering approach makes it possible to isolate these signals and determine their temporal distributions, which is another first step toward interpreting them.
The precise determination of temporal distribution and the discovery of unknown signals together demonstrate the effective exploratory capabilities offered by this method.

Compared to existing self-supervised clustering approaches that operate on whole-window embeddings, the proposed pipeline explicitly extracts event-level representations within each spectrogram before performing dataset-level clustering. This distinction is particularly relevant for hydroacoustic data, where multiple sources often overlap in time and frequency. A single embedding vector representing an entire time window can therefore conflate several concurrent signals. By contrast, clustering patch embeddings within each spectrogram enables the identification of individual events and the assignment of separate embedding vectors to each of them. This intermediate step makes it possible to disentangle coexisting sources, such as whale calls occurring in the presence of ship noise, and improves the interpretability of the resulting clusters. However, the proposed approach fails to disentangle time-frequency overlapping signals in spectrograms. Still, Table \ref{table_eval} shows that the proposed method yields better results than the proposed spectrogram-level embedding approaches. In particular, the technique of averaging the patch embeddings performs poorly, which is expected as most patches represent noise. Similarly, the technique of selecting the embedding with the highest norm also largely fails. Given that the technique of selecting the embedding vector being the most distant from the spectrogram-level average performs well, this suggests that the noisy components of the spectrogram may have embedding vectors with a norm significantly greater than zero, leading to a simple maximum-looking method to select noisy patches. Notably, one of those techniques enabled a higher result for EQ\_P than event extraction techniques, likely caused by earthquake signals being responsible for most of the energy of their time window. Further experimentation including contrastive learning is required to obtain a fair comparison with contrastive-based techniques. In particular, more advanced methods relying on embedding space constraints \citep[e.g. DINO,][]{caron_emerging_2021} may lead to improvements.

Given that clustering represents one of the main stages of the method, a comparison of different clustering techniques is worthwhile. Table \ref{table_eval} shows \textit{MAE-K-Means} outperforms \textit{MAE-UMAP} in most cases. This may be due to the space distortion caused by UMAP. Moreover, the dimension reduction of UMAP may cause a loss of information, while K-Means was directly run on the original embedding space. Two problems were expected for K-Means: first, while the UMAP-distortion is avoided, running K-Means on the original embedding space may suffer a loss of euclidean distance relevance due to high-dimensional distance concentration. Then, a problem of K-Means is that it clusters all data points while HDBSCAN explicitly discards regions of space which are not dense enough, classifying them as noise. The former problem could make cosine-distance-based techniques good alternatives. Despite this, K-Means performs better, suggesting the problems induced by UMAP outweigh the distance-concentration effect. The latter problem may explain the higher cluster sizes as compared to \textit{MAE-UMAP}. In particular, the highest cluster size exceeds 300,000. However, this cluster was associated with noise as expected, and as stated in the caption of Table \ref{table_classes}, the majority of events were effectively classified as noise, which is expected. Lastly, replacing UMAP with PCA gave poor reasults. This expected behavior may be attributed to the linear nature of PCA, which poorly captures the structure of the data.

The ablation of pre-training led to a model that performs from lower to significantly lower than the pre-trained \textit{MAE-K-MEANS}, except for SWIO-PBW where it obtains good results. These results demonstrate that while pre-training is not strictly required for performance on large-scale datasets, it remains a critical factor for achieving optimal clustering accuracy, thereby confirming its role in enhancing feature representation.

Several limitations and possible extensions should be considered. First, the sensitivity of the pipeline to analysis parameters such as window duration, spectrogram resolution, spectrogram normalization or clustering hyperparameters was not explored extensively. Similarly, we did not try different models of MAE or different pre-training datasets, including non-audio datasets. This is largely due to the training time of the MAE which, while still acceptable using a personal computer, is too long to iterate the training phase with different setups. However, preliminary tests showed limited sensitivity within the reasonable ranges of parameters, in particular regarding HDBSCAN or spectrogram resolution. Future work could investigate lighter or, conversely, larger architectures or transfer-learning strategies to reduce the training cost and ease experimentation. Secondly, the mapping from clusters to semantic classes currently relies on manual inspection. Although this step requires relatively little time compared to a full annotation of the dataset, semi-automatic or active-learning approaches could further streamline the process. Finally, while the method was applied here to data sampled at 240\,Hz, it is in principle applicable to higher-frequency recordings and to other types of passive acoustic monitoring (PAM) including aerial PAM, provided that appropriate spectrogram parameters are chosen.

\section{Conclusion}
The results show that a self-supervised clustering strategy at the event level can be a powerful tool for exploring large low-frequency hydroacoustic datasets with minimal annotation effort. By combining representation learning, within-spectrogram event extraction, and dataset-level clustering, the proposed pipeline yields clusters with clear semantic meaning and enables rapid construction of long-term activity time series. Such capabilities prove particularly useful in situations where labeled data is scarce but large volumes of recordings are available. Future developments of this or other self-supervised methods can be quantitatively compared using the classification-based evaluation adopted here.
Lastly, as passive hydroacoustic monitoring networks continue to operate around the world, these types of methods should greatly facilitate the extraction of first-order information from unexplored archives, whether through the discovery of unknown signals or the assessment of temporal distributions.

\section*{Author Declarations}
The authors declare no conflicts of interest.

\section*{Data Availability}
An open-source version of the code together with the pre-trained MAE weights used in this work are available on Github \citep{zenodo_code}. The MAHY data used as the evaluation dataset are available as FDSN network 1T (DOI 10.15778/RESIF.1T2018).

\begin{acknowledgments}
We thank the captains and crews of R/V Marion Dufresne and Pourquoi Pas ? for the successful deployments and recoveries of the autonomous hydrophones used in this work. We also thank the CTBTO preparatory commission for giving access to IMS data for scientific purposes. The commission is not responsible for the views of the author. This work is part of the project MUSIC funded by ANR (ANR-23-EDIR-0002). At the start of this study, Pierre-Yves Raumer was supported by a PhD fellowship from the University of Brest and from the Regional Council of Brittany (ARED), through the Interdisciplinary Graduate School for the Blue Planet (ISblue), co-funded by ANR (ANR-17-EURE-0015). He was then funded by ERC SeaSALT (ERC grant SeaSALT $\#$101170619). Lastly, we would like to thank the Datarmor team for providing and maintaining the Datarmor cluster, which was used in the initial pre-training phase of this work.

\end{acknowledgments}

\section*{References}

\bibliography{zotero.bib, external.bib}

\end{document}